\documentclass[12pt]{article}
\begin{document}
%%%%%%%%%%%%%%%%%%%%%%%%%%%%%%%%%%%%%%%%%%%%%%%%%%%%%%%%%%%%%%%%%%%%%
\title{A gauge invariant formulation for the $SU(N)$ non-linear
  $\sigma$-model in $2+1$ dimensions}
\author{C.\ D.\ Fosco$^a$\thanks{CONICET}\\
  and\\
  C.\ P.\ Constantinidis$^b$\thanks{CNPq}\\
  {\normalsize\it $^a$Department of Physics, Theoretical Physics,}\\
  {\normalsize\it 1 Keble Rd., Oxford OX1 3NP, United Kingdom}\\
  {\normalsize\it $^b$Universidade Federal do Esp\'{\i}rito Santo,}\\
  {\normalsize\it 29060-900 Vit\'oria - ES, Brasil}} \date{\today}
\maketitle
%====================================================================
\begin{abstract}
\noindent
We derive a local, gauge invariant action for the $SU(N)$ non-linear
$\sigma$-model in $2+1$ dimensions. In this setting, the model is
defined in terms of a self-interacting pseudo-vector field
$\theta_\mu$, with values in the Lie algebra of the group $SU(N)$.
Thanks to a non-trivially realized gauge invariance, the model has the
correct number of physical degrees of freedom: only one polarization
of $\theta_\mu$, like in the case of the familiar Yang-Mills theory in
$2+1$ dimensions.  Moreover, since $\theta_\mu$ is a pseudo-vector,
the physical content corresponds to one massless {\em pseudo-scalar\/}
field in the Lie algebra of $SU(N)$, as in the standard representation
of the model.  We show that the dynamics of the physical polarization
corresponds to that of the $SU(N)$ non-linear $\sigma$-model in the
standard representation, and also construct the corresponding BRST
invariant gauge-fixed action.
\end{abstract}
\bigskip
%====================================================================
\section{Introduction}\label{sec:intro}
The non-linear $\sigma$-model~\cite{weinberg1} is a very important
tool for the description of the effective, low-energy dynamics of
systems with a broken continuous (global) symmetry~\cite{zinn}. Many
of its interesting and distinctive features stem from the fact that
the symmetry group is realized in a non-linear way, as this endows the
theory with a rich structure of interactions. Indeed, it has an
infinite number of interaction vertices, when defined in terms of
field variables which are themselves group coordinates.  Nonetheless,
this holds true in spite of the model having a `universality': its
properties are completely determined when the symmetry group and the
spacetime dimension are known.

Of course, the same non-linearity is also responsible for the fact
that, except for the $1+1$ dimensional case, the theory becomes
non-renormalizable from the point of view of the usual loop
expansion~\cite{zinn}.  However, even in more than two spacetime
dimensions, the model still has a reasonable predictive power, if
properly understood as an effective theory~\cite{weinberg2}.  This
approach has been successfully applied to chiral perturbation
theory~\cite{cpt}, as a convenient effective model for $QCD$.  Note,
however, that in $2+1$ dimensions, the non-linear $\sigma$-model is
renormalizable if a large-$N$ expansion is used~\cite{largen}, instead
of the standard loopwise perturbation theory.

The non-linearity may usually be tackled by resorting to an auxiliary,
`Lagrange multiplier' field, which enforces a constraint on the
(otherwise free) field variables. The typical example of this is,
perhaps, the $O(N)$ non-linear $\sigma$-model, where an auxiliary
field imposes a constant-modulus constraint on an $N$-component scalar
field ${\vec \phi}\,=\,(\phi_1,\ldots,\phi_N)$, which is a vector
field in internal space.  An important by-product of this construction
is that the auxiliary field is a $O(N)$ singlet, hence, the large-$N$
expansion is easier to formulate after one `integrates out' the $\phi$
field, leaving an action for the Lagrange multiplier.

Indeed, the procedure of `linearizing' an action, by the introduction
of auxiliary fields, and afterwards integrating the original fields
out to obtain an effective theory for the auxiliary fields, has
frequently proven to be very useful. This is particularly true when
the auxiliary field has some convenient symmetry or transformation
properties~\cite{rivers}.  In particular, it allows one to obtain an
effective theory where the symmetry properties are inherited from the
ones of the Lagrange multiplier in the linearized theory.

In this paper, we introduce a gauge invariant, non-trivially realized
Abelian quantum field theory model in $2+1$ dimensions, which is
derived by the procedure of integrating out the original variables, in
order to obtain an effective theory for the auxiliary field. Since our
starting point shall be a representation of the non linear
$\sigma$-model where the Lagrange multiplier has a local gauge
symmetry, that feature will be preserved in the resulting action.  The
realization of the Abelian gauge symmetry is non trivial, because the
commutator of two gauge transformations is zero only on-shell, i.e.,
on the configurations that satisfy the equations of motion.
Equivalently, the commutator between two `true' gauge transformations
yields a trivial, `equation of motion' gauge
transformation~\cite{deWit:1978cd,teit}.

The structure of this paper is as follows: in
section~\ref{sec:themodel} we derive the action for model, showing
that it is indeed defined by a gauge invariant action. Then we
consider the realization and structure of the gauge and global
symmetries in section~\ref{sec:gauge}, leaving for
section~\ref{sec:quantum} the quantum treatment of the model.
Section~\ref{sec:conc} contains our conclusions.

\section{The model}\label{sec:themodel}
We shall begin by reviewing the main features of the polynomial
representation for the $SU(N)$ non-linear $\sigma$-model in $2+1$
dimensions, as presented in~\cite{fm1,fm2}.  This formulation may be
defined in terms of a gauge invariant Euclidean action $S_{inv}$,
which determines the dynamics of two fields $L_\mu$ (vector) and
$\theta_\mu$ (pseudo-vector) in the Lie algebra of $SU(N)$:
\begin{equation}\label{eq:defseuc}
S_{inv}[L,\theta] \;=\; \int d^3x \,{\mathcal L}_{inv} (L,\theta)
\end{equation}
with
\begin{equation}\label{eq:defleuc}
{\mathcal L}_{inv}(L,\theta)\;=\; \frac{1}{2} g^2 L_{\mu}\cdot L_\mu
+ i g \,\theta_\mu \cdot {\tilde F}_\mu(L)
\end{equation}
where $g$ is a constant with the dimensions of a mass (it is in fact
the exact analog of $f_\pi$ in the $3+1$ dimensional case), and
${\tilde F}_\mu (L)$ denotes the dual of the non Abelian field
strength tensor for the vector field $L_\mu$, namely,
\begin{equation}\label{eq:defFt}
{\tilde F}_\mu (L) \;=\; \frac{1}{2}\epsilon_{\mu\nu\lambda} F_{\nu\lambda}(L)
\;\;\;,\;\;\;
F_{\mu\nu}(L) \;=\; \partial_\mu L_\nu - \partial_\nu L_\mu
\,+\, g^{\frac{1}{2}} [L_\mu , L_\nu ] \;.
\end{equation}
Being $L_\mu$ an element in the Lie algebra, with the convention that
$L_\mu = - L_\mu^\dagger$, it can be written as
$$
L_\mu(x) \;=\; L_\mu^a (x) \lambda_a \;\;,\;\; \lambda_a^\dagger
\;=\; - \lambda_a \;,
$$
\begin{equation}\label{eq:conv}
{\rm tr} (\lambda_a \lambda_b) \;=\; - \delta_{a b}\;\;,\;\;
[\lambda_a , \lambda_b ] \;=\; f_{a b c} \, \lambda_c
\end{equation}
where $f_{a b c}$ is real and completely antisymmetric. Group indices
will be indistinctly written as subscripts or superscripts; no meaning
should be assigned to the difference.  In (\ref{eq:defleuc}), we also
used the notation: $U \cdot V \equiv U_a V_a$, and $(U \times V)_a =
f_{a b c} U_b V_c$ for any two elements $U$, $V$ in the algebra. Also,
both $L$ and $\theta$ have the mass dimensions of $g^{1/2}$.

The `inv' subscript in the action has been introduced in order to
emphasize the fact that it is, indeed, invariant under the (local)
gauge transformations:
\begin{equation}\label{eq:pgtrns}
\delta_\omega L_\mu \;=\; 0 \;\;\;\;\;\;\;
\delta_\omega \theta_\mu \;=\; D_\mu \omega \;,
\end{equation}
where the covariant derivative is compatible with the parallel
transport defined by $L$, namely,
\begin{equation}
D_\mu \omega = \partial_\mu \omega + g^{\frac{1}{2}} [L_\mu , \omega]
\;,
\end{equation}
or in components:
\begin{equation}
(D_\mu \omega)^a = \partial_\mu \omega^a + g^{\frac{1}{2}}
\, f_{a b c} \, L_\mu^b \,  \omega_c \;.
\end{equation}

It must be noted that this gauge symmetry is valid {\em of-shell},
namely, it holds true regardless of whether the fields verify the
equations of motion or not. Besides, equation (\ref{eq:pgtrns}) tells
us that $L$ is a gauge-invariant object, and this implies that the
commutator of two gauge transformations vanishes:
\begin{equation}\label{eq:polcomm}
\left[\delta_\eta \,,\, \delta_\omega \right]\;=\; 0 \;.
\end{equation}
Here $\delta_\omega$ and $\delta_\eta$ denote the operators that
perform a gauge transformation on a given functional (eventually a
function) of the fields.  Namely, if $I$ is a functional of $L$ and
$\theta$,
\begin{equation}
\delta_\omega I[L,\theta] \,=\, \int d^3x \, \delta_\omega
\theta_\mu^a (x) \, \frac{\delta I[L,\theta]}{\delta \theta_\mu^a (x)} 
\;,
\end{equation}
where $\delta_\omega \theta_\mu^a$ is defined as in (\ref{eq:pgtrns}).
This of course means that the gauge group is {\em Abelian}, in spite
of the non-Abelian looking transformation rule for $\theta$.

Had we wanted to work with this representation, we should have
considered fixing the gauge as the next step. Rather than doing that,
we shall move on to derive an `effective theory' for $\theta_\mu$, an
auxiliary field which transforms as a vector field in the adjoint
representation.  To that end, we define the effective action
$S_{inv}[\theta]$ by the following expression:
\begin{equation}\label{eq:defst}
\int [{\mathcal D} \theta ] \; e^{ - S_{inv}[\theta] }
\;=\; \int {\mathcal D}\theta \, {\mathcal D}L \; e^{-S_{inv}[L,\theta]}
\end{equation}
where $[{\mathcal D}\theta]$ denotes the integration measure for
$\theta$ in the effective theory (the brackets denote possible group
factors).  Of course, the integration over $\theta_\mu$ is
ill-defined, since the theory is gauge invariant. There is, however,
no obstruction to the integration of the $L$-field, since $\theta_\mu$
is, in that case, regarded as a background field.  We shall, of
course, have to deal with the gauge-fixing for $S_{inv}[\theta]$
afterwards.

The integral over $L_\mu$ in (\ref{eq:defst}) is a Gaussian, and its
evaluation yields the result:
\begin{equation}\label{eq:st}
S_{inv}[\theta] \;=\; \int d^3x \, {\mathcal L}_{inv}(\theta)\;\;,\;\;
{\mathcal L}_{inv}(\theta) \;=\; \frac{1}{2} {\tilde f}_\mu^a
G_{\mu\nu}^{ab}(\theta){\tilde f}_\nu^b
\end{equation}
where $\tilde f$ is the dual of the {\em Abelian\/} field~\footnote{
  We adopt the convention that a lowercase $f_\mu$ refers to the dual
  of the {\em Abelian\/}field strength, while the uppercase one is
  reserved for the dual non Abelian one.} strength: \mbox{${\tilde
    f}_\mu^a \equiv \epsilon_{\mu\nu\lambda}\partial_\nu
  \theta^a_\lambda$}, and
\begin{equation}\label{eq:defg}
G_{\mu\nu}^{ab} \,=\, [ M^{-1} ]_{\mu\nu}^{ab}
\;\;,\;\;
M_{\mu\nu}^{ab}\,=\, \delta_{\mu\nu}\delta^{ab} + i g^{-\frac{1}{2}}
\epsilon_{\mu\lambda\nu} f^{acb} \theta_\lambda^c\;.
\end{equation}
The fact that $G$ is the inverse of $M$ must be understood in the
sense that the relations:
\begin{equation}\label{eq:definv}
G_{\mu\lambda}^{ac} M_{\lambda\nu}^{cb} \;=\; \delta_{\mu\nu} \delta^{ab}
\end{equation}
are valid. Fortunately, the explicit for of $G$ is not required for
most of our presentation. Note, however, that one may easily obtain an
approximate expression for it by performing an expansion in powers of
the (dimensionless) object $\theta g^{-\frac{1}{2}}$.  There arises
also from the Gaussian integral a factor which modifies the
$\theta$-field integration measure,
\begin{equation}\label{eq:thmeas}
[{\mathcal D}\theta]\;=\; {\mathcal D}\theta \, [\det(M)]^{-\frac{1}{2}}
\end{equation}

A question that immediately presents itself at this point is what has
happened to the gauge invariance; indeed, the gauge invariance in the
polynomial representation, equation (\ref{eq:pgtrns}), involves
$L_\mu$ in its definition, and $L_\mu$ is precisely the field that has
been eliminated from the action.

Of course, a standard Maxwell-like gauge transformation will not do,
since, although ${\tilde f}_\mu$ is invariant under the Abelian gauge
transformations of the Maxwell theory, $G$, that depends on
$\theta_\mu$, is not. Indeed, looking for example at the explicit form
of the action (\ref{eq:st}), with $G$ expanded up to terms of order
$\frac{\theta^2}{g}$, we see that:
$$
S_{inv}[\theta] \;=\; \int d^3x \, \left[ \frac{1}{2} {\tilde
    f}_\mu(\theta) \cdot {\tilde f}_\mu(\theta) \,-\,\frac{i}{2}
  g^{-\frac{1}{2}} \epsilon_{\mu\nu\lambda} \, \theta_\mu \cdot
  {\tilde f}_\nu (\theta) \times {\tilde f}_\lambda(\theta) \right.
$$
\begin{equation}
\left. - \frac{1}{2 g} ( \theta_\mu\cdot{\tilde f}_\mu  
\theta_\nu\cdot{\tilde f}_\nu  \,-\,
\theta_\mu\cdot{\tilde f}_\nu \,  \theta_\mu\cdot{\tilde f}_\nu \,
+\, {\tilde f}_\mu\cdot{\tilde f}_\mu \,  \theta_\nu\cdot \theta_\nu
  -\, {\tilde f}_\mu\cdot{\tilde f}_\nu  \,  \theta_\mu \cdot
  \theta_\nu )
\right] \;,
\end{equation}
where only the term in the first line is invariant under Abelian gauge
transformations.  In spite of this, we do expect a gauge invariance to
exist for $S_{inv}[\theta]$, since we know there are two unphysical
components (for each value of $a$) in $\theta_\mu$, which do appear in
the free propagator. This propagator will of course be determined by
the free action
\begin{equation}\label{eq:mxwll}
S^{(0)}_{inv} [\theta]\;=\; \int d^3x\, \frac{1}{2} {\tilde f}_\mu^a(\theta)
{\tilde f}_\mu^a(\theta) \;=\; \int d^3x \;\frac{1}{4} f^a_{\mu\nu}(\theta) f^a_{\mu\nu}(\theta)
\end{equation}
after adding a gauge fixing term.

It is then reasonable to assume that the gauge transformations for
$\theta$ should be of the form
\begin{equation}\label{eq:egtrns}
\delta_\omega \theta_\mu \;=\; \partial_\mu \omega + g^{\frac{1}{2}} [L_\mu(\theta) , \omega]
\end{equation}
where $L_\mu(\theta)$ is a {\em dependent\/} field which plays the
role of a connection, and should of course be defined in terms of
$\theta$.

A possible hint to find the explicit form of $L_\mu(\theta)$ comes
from the fact that performing the Gaussian integration is tantamount
to `replacing the integrated field by their values at the extreme of
the exponent'.  Denoting by ${\hat L}_\mu(\theta)$ the expression that
maximizes the exponent, we see that it is given by:
\begin{equation}\label{eq:ltheta}
{\hat L}^a_\mu \;=\; - i g^{-1} \, G_{\mu\nu}^{ab}(\theta) {\tilde f}_\nu^b\;.
\end{equation}
Thus we shall adopt the ansatz $L_\mu(\theta) \equiv {\hat L}_\mu
(\theta)$, the consistency of which we will verify now: to see whether
the transformation (\ref{eq:egtrns}) is a (gauge) symmetry of the
action (\ref{eq:st}) or not, we first evaluate the first variation of
$S_{inv}[\theta]$ under a general, not necessarily gauge,
infinitesimal variation of $\theta$.

After some elementary algebra, we obtain:
$$
\delta S_{inv}[\theta] \;=\; \int d^3x \,\delta\theta_\mu^a \,
\left\{ \epsilon_{\mu\nu\lambda}\,\partial_\nu
[  G_{\lambda\rho}^{ab}(\theta) {\tilde f}_\rho^b(\theta) ] \right.
$$
\begin{equation}\label{eq:fvar}
\left. -\frac{i}{2} g^{-\frac{1}{2}} \,  \epsilon_{\mu\nu\lambda}
\, f_{abc}\, G_{\nu\alpha}^{bd}(\theta)\,{\tilde f}_\alpha]^d
\, G_{\lambda\beta}^{ce}(\theta) \, {\tilde f}_\beta^e \right\}
\end{equation}
where we used the symmetry property $G_{\mu\nu}^{ab} = G_{\nu\mu}^{ba}$, and
the relation
\begin{equation}
\delta G_{\mu\nu}^{ab} \;=\; - i g^{-\frac{1}{2}}\,
G_{\mu\lambda}^{ac}(\theta)\epsilon_{\lambda\rho\sigma} \, f^{cde} \,
\delta\theta_{\rho}^d \,  G_{\sigma\nu}^{eb} (\theta)\;,
\end{equation}
both of them consequences of the fact that $G = M^{-1}$.
Recalling the definition of $L_\mu(\theta)$, we may also write
(\ref{eq:fvar}) as:
\begin{equation}\label{eq:fvar1}
\delta S_{inv}[\theta] \;=\; i g \int d^3x \, \delta\theta_\mu^a \,
{\tilde F}_\mu^a (L(\theta))
\end{equation}
where
$$
{\tilde F}_\mu^a (L(\theta)) \;=\; \frac{1}{2}
\epsilon_{\mu\nu\lambda} F^a_{\nu\lambda}(L(\theta)) \;,
$$
\begin{equation}\label{eq:defF}
F_{\mu\nu}^a (L(\theta)) \;=\; \partial_\mu L_\nu^a (\theta) - \partial_\nu
L_\mu^a(\theta) + g^{\frac{1}{2}} f^{abc}L_\mu^b(\theta) L_\nu^c
(\theta)\;.
\end{equation}
Using now the explicit form for $\delta\theta_\mu$ that corresponds to
a gauge variation, equation (\ref{eq:egtrns}), we see that
\begin{equation}
\delta S_{inv}[\theta] \;=\; - i g \int d^3x \; \omega^a (x)
[D_\mu {\tilde F}_\mu ]^a(L) \;=\;0 \;,
\end{equation}
as a consequence of the Bianchi identity, which is of course true
regardless of $L$ being an independent field or not. We shall
henceforth omit writing the dependence of $L$ on $\theta$ explicitly,
since $L$ shall always be assumed to be a dependent field.  A small
technical point (absent in the real time formulation) is that the
relation (\ref{eq:ltheta}) includes complex factors: an $i$
multiplying $G$, but also $G$ itself has both real an imaginary parts.
That should be hardly surprising, since the action itself is not
purely real, as it happens with Euclidean actions including
Chern-Simons terms (and with other topological objects in different
numbers of dimensions).  Thus the relation (\ref{eq:ltheta}), to have
non-trivial solutions, require the continuation of the fields to
complex values.  Of course, the gauge invariant action in Minkowski
spacetime, $S_{inv}^M$, is real,
\begin{equation}\label{eq:mink}
S_{inv}^M \,=\, \int d^3x \,\frac{1}{2} {\tilde f}^\mu_a
G_{\mu\nu}^{ab}(\theta){\tilde f}^\nu_b
\end{equation}
where ${\tilde f}^\mu_a = \epsilon^{\mu\nu\lambda}\partial_\nu
\theta_\lambda^a$ and $G_{\mu\nu}^{ab}(\theta)$ is determined by
the equations:
\begin{equation}\label{eq:defgm}
G_{\mu\rho}^{ac}(\theta)\, M^{\rho\nu}_{cb}(\theta) \;=\;
\delta_\mu^\nu \delta^a_b 
\;\;,\;\;
M^{\mu\nu}_{ab}\,=\, g^{\mu\nu}\delta_{ab} +  g^{-\frac{1}{2}}
\epsilon^{\mu\lambda\nu} f^{acb} \theta_\lambda^c\;.
\end{equation}

Thus we have verified the consistency of the definition of the
covariant derivative with the gauge invariance of the action. Note,
however, that there is an important difference with the polynomial
formulation, in that the gauge transformations for $\theta$ involve
$L$, which is itself a function on $\theta$.  Thus $L$ will, in
general, change under a gauge transformation in this formulation. In
particular, this implies that finite gauge transformations will be
different to infinitesimal ones. This is a consequence in fact of the
algebra of gauge transformations being open, as it will be discussed
in the next section.

Also, expression (\ref{eq:fvar1}) tells us that the classical
equations of motion deriving from $S_{inv}[\theta]$ are:
\begin{equation}\label{eq:mc}
F_{\mu\nu}(L) \;=\; 0 \;.
\end{equation}
i.e., the Maurer-Cartan equations for $L$, which obviously have a
gauge invariant set of solutions.  

Regarding the integration measure $[{\mathcal D}\theta]$, it is
straightforward to verify that the gauge variation of $[{\mathcal
  D}\theta]$ is zero.  We conclude that the action (\ref{eq:st}) is
indeed gauge invariant.  The gauge invariance is not of the Yang-Mills
type, but rather involves as a connection a vector field $L_\mu$ which
is a composite field, defined in terms of $\theta_\mu$ and its
derivatives.  As we shall see in the next section, the gauge group is
indeed Abelian, but the albegra of gauge transformations is not closed
off-shell.
 
It may seem surprising at first sight that the only `content' of the
classical equations of motion is that the Maurer-Cartan equations for
a field are satisfied, since we still need the dynamics for the true
degrees of freedom. Of course, such a dynamics is also present in this
description: $L$ is a pure gauge field, i.e., $L_\mu = U^\dagger
\partial_\mu U$ with $U(x) \in SU(N)$, and besides (see
(\ref{eq:global}) below) $\partial_\mu \cdot L_\mu = 0$. These two
equations are the equivalent to the classical equations of motion for
the non-linear $\sigma$-model.

\section{Symmetries}\label{sec:gauge}
The actual form of the gauge transformations, as acting on the field
$\theta_\mu$, has been obtained by the procedure of borrowing the
(known) form of the corresponding transformations from the polynomial
version, and afterwards replacing the field $L_\mu$ by its value at
the extreme (a function of $\theta$). This yields, for a
transformation parametrized by the function $\omega(x)$, the
variation:
\begin{equation}\label{eq:vartheta}
\delta_\omega \theta_\mu(x) \;=\; D^L_\mu \omega(x)
\end{equation}
where
\begin{equation}\label{eq:defdl}
D^L_\mu \omega \;=\; \partial_\mu \omega + g^{\frac{1}{2}} [L_\mu,\omega] \;,  
\end{equation}
with: 
\begin{equation}\label{eq:loft}
L_\mu^a \,=\, -i g^{-1} \, G_{\mu\nu}^{ab}(\theta)\, {\tilde f}_\nu^b (\theta)\;.
\end{equation}
In spite of the presence of a covariant derivative, the
transformations do not correspond to a non-Abelian Yang-Mills theory.
Indeed, it should be noted that the transformations
(\ref{eq:vartheta}) involve the covariant derivative, defined in terms
of a composite field which plays the role of a connection.  However,
they are not of the strictly Abelian type either, since the
transformation law for $\theta$ does not correspond to that case.

We shall now see that what happens is that the transformations  are, 
indeed, Abelian, but only {\em on-shell}, i.e.,
on the equations of motion. 
To be specific, consider the commutator of two gauge transformations, 
corresponding to the gauge functions $\omega$ and $\eta$.  We find that the 
result may be written, after some algebraic manipulations, as follows:
\begin{equation}\label{eq:gaugecomm}
[\delta_\eta, \delta_\omega ] \theta_\mu^a \;=\; 
\Sigma_{\mu\nu}^{ab} (\theta) \, \frac{\delta S_{inv}[\theta]}{\delta \theta_\nu^b}
\end{equation}
where we introduced the object:
\begin{equation}\label{eq:defsigma}
\Sigma_{\mu\nu}^{ab}(\theta) \;=\; - \frac{1}{g} \,
\eta^h \, \omega^c \, ( f^{aec}   f^{dbh} -   f^{aeh}   f^{dbc})
G_{\mu\nu}^{ed}(\theta) \;. 
\end{equation}
It is important to realize that $\Sigma_{\mu\nu}^{ab}$ is antisymmetric,
namely,
\begin{equation}\label{eq:asymm}
\Sigma_{\mu\nu}^{ab} \;=\; -\Sigma_{\nu\mu}^{ba} \;, 
\end{equation}
since this means that the right hand side of (\ref{eq:gaugecomm}) is a
trivial gauge transformation~\cite{teit}.
Indeed, for a given action $S[\theta]$, a transformation of the kind
\begin{equation}
\delta \theta_\mu^a \;=\; \Lambda_{\mu\nu}^{ab}(\theta) \frac{\delta
S[\theta]}{\delta \theta_\nu^b} 
\end{equation}
with an arbitrary antisymmetric function $\Lambda_{\mu\nu}^{ab}=-
\Lambda_{\nu\mu}^{ba}$, is a symmetry of $S[\theta]$, regardless  of the 
form of $S[\theta]$. It can also be shown~\cite{teit}, that the commutator 
between a non-trivial gauge transformation and a trivial one yields a trivial 
gauge transformation.
Thus, we see that the physically relevant gauge group is Abelian, and
isomorphic to $U(1)^{(N^2 -1)}$ (for $SU(N)$), although realized in a non-trivial 
way, since the `trivial' part of the gauge transformations cannot be
easily eliminated within the present formulation of the model. 

A related property is that the composite field $L_\mu$, which is gauge invariant 
in the polynomial transformation, is now also gauge-invariant but only on-shell:
\begin{equation}\label{eq:delf}
\delta_\omega L_\mu^a \;=\; - i g^{-\frac{1}{2}} \, G_{\mu\nu}^{ab}(\theta) 
f^{bcd} {\tilde F}_\nu^c (L) \omega_d  \;,
\end{equation}
i.e., it vanishes when ${\tilde F}_\mu (L)= 0$.

%%%%%%%%%%%%%%%%%%%%%%%%%%%%%%%%%%% Algebra of the Gauge Group %%%%%%%%%%%%%%%%%%%%%%%%%%%%%%%%%%%%%%%%
The question that immediately presents itself is what are the
conditions a gauge invariant functional must verify. This is of course
important, since gauge invariant functionals are naturally associated
to physical observables. Besides, in the functional integral approach
to a quantum gauge field theory, the condition a gauge invariant
functional must satisfy is an important part of the formulation.

So, assuming $I[\theta]$ to be a gauge invariant
functional of $\theta$, it must verify the condition:
\begin{equation}\label{eq:ginvi}
\delta_\omega \, I[\theta] \;=\; 0 \;,
\end{equation}
where 
\begin{equation}
\delta_\omega \;=\; \int d^3x \, \delta_\omega \theta_\mu^a(x) \, 
\frac{\delta}{\delta \theta_\mu^a(x)} \;.
\end{equation}
However, if such a gauge invariant functional exists, one immediately
gets a consistency condition by applying two successive gauge
transformations on $I$ and subtracting them, namely:
\begin{equation}\label{eq:ccond}
\delta_\omega I[\theta]=0 \;\;\Rightarrow \;\; 
\left[ \delta_\eta \,,\, \delta_\omega \right] I[\theta] \,=\, 0 \;.
\end{equation}
On the other hand, we may of course evaluate the commutator of two
gauge transformations; after some algebra, we find:
\begin{equation}\label{eq:comm}
[ \delta_\eta \,,\, \delta_\omega ] \;=\; \int d^3x \,
\Sigma_{\mu\nu}^{ab}(\theta) \, \frac{\delta S_{inv}}{\delta
  \theta_\mu^a (x)}  \frac{\delta }{\delta\theta_\nu^b (x)}  \;.
\end{equation}
Thus, for non-trivial gauge invariant functional $I$ to exist, since
$\Sigma$ depends on the arbitrary functions $\eta$ and $\omega$, we
have to impose the additional condition:
\begin{equation}\label{eq:extra}
F_{\mu\nu}(L) \;=\; 0\;.
\end{equation}
This is nothing new from the classical point of view, but it makes a
difference for the quantum theory, where all the configurations
matter, and not just the extrema of the action. This seems to lead us
to the inclusion of (\ref{eq:extra}) as a constraint, what is not what
we want. Fortunately, there are ways out of this~\cite{teit}, that
does not require the introduction of  extra constraints (wich might
even reduce the number of degrees of freedom. 

%%%%%%%%%%%%%%%%%%%%%%%%%%%%%%%%%%%%%%%%%%%%%%%%%%%%%%%%%%%%%%%%%%%%%%%%%%%%%%%%%%%%%%%%%%%%%%%%%%%%%
Regarding the global symmetries, we know that $L_\mu$ is a conserved current, 
associated to a global symmetry of the non-linear $\sigma$-model. To see that 
$L_\mu$ is conserved in this formalism is a bit tricky. One possible way to
prove that is to use the property that the composite field $L_\mu$ as given 
by (\ref{eq:loft}) may also be written, after some algebra, as:
\begin{equation}\label{eq:eqlt}
L_\mu \;=\; -i g^{-1} \, \epsilon_{\mu\nu\lambda} D_\nu \theta_\lambda \;,
\end{equation}
where we used the property:
\begin{equation}
G_{\mu\nu}^{ab} (\theta) \;=\; \delta_{\mu\nu}^{ab} \,-\, i g^{-\frac{1}{2}}
\epsilon_{\mu\lambda\sigma} f^{acd} \theta_\lambda^c G_{\sigma\nu}^{db}(\theta)
\;. 
\end{equation}
Then it follows that
\begin{equation}\label{eq:global}
\partial_\mu L_\mu \;=\; D_\mu L_\mu \;=\; -i g^{-1} \, \epsilon_{\mu\nu\lambda} 
D_\mu D_\nu \theta_\lambda \;=\;  -i g^{-\frac{1}{2}} [ {\tilde F}_\mu (L),
\theta_\mu ]  
\end{equation}
which vanishes on shell, and implies the conservation of $L_\mu$.
The conserved charge is of course given by the space integral of $L_0$. It is 
instructive to consider the particular case of a static point-like static charge
of color $a$ and strength $q$ located at ${\mathbf x} = {\mathbf x}_0$. 
This corresponds to a charge density
\begin{equation}
L^a_0 (x) \;=\; - i q \delta ({\mathbf x} - {\mathbf x}_0) ,\;\;\;
L_j (x) \;=\; 0\;.
\end{equation}
Inserting this into the relation (\ref{eq:loft}) yields
\begin{equation}
{\tilde f}_\mu^a \;=\; q \, \delta_{\mu 0} \, 
\delta({\mathbf x} - {\mathbf x}_0)
\end{equation}
i.e., it corresponds to a point like magnetic flux sitting on the same point.
The conserved charge is then equal to the total magnetic flux (for that color).

\section{Quantum theory}\label{sec:quantum}
We shall consider here the quantum theory corresponding to this gauge
invariant model, from the path integral approach. The natural object to
consider is then of course the generating functional for $\theta$-field
correlation functions. The ill-defined (gauge invariant) partition function
shall be given by the expression:
\begin{equation}\label{eq:defzinv}
{\mathcal Z}_{inv}[J] \;=\; \int [{\mathcal D}\theta] \,
\,\exp \left\{- S_{inv}[\theta] + \int d^3x J_\mu \cdot \theta_\mu \right\} \;.
\end{equation}

The generating functional (\ref{eq:defzinv}), being gauge invariant,
requires the introduction of a gauge-fixing term and its companion
ghost action to be well-defined. However, a standard Faddeev-Popov
approach to the definition of the gauge-fixed action will not do,
since the resulting action is neither BRST invariant, nor the
transformation becomes nilpotent.  The difficulty lies, of course, in
the fact that the algebra of the gauge transformations is `open',
namely, it closes only when the equations of motion are satisfied.
However, a modified action, which generally involves quartic ghost
terms may be constructed, such that the action is invariant under an
extended BRST transformation~\cite{deWit:1978cd,teit}.  By an application
of such method to this case, we obtain the gauge-fixed action $S$:
\begin{equation}\label{eq:defsfx}
S[\theta_\mu;b,{\bar c}, c] \;=\; S_{inv}[\theta] \,+\,
S_{gf}[b,\theta]\,+\, S_{gh}[{\bar c},c;\theta] 
\end{equation}
where 
We shall adopt the covariant gauge-fixing term:
\begin{equation}\label{eq:defsgf}
S_{gf}[\theta] \;=\; \int d^3x \,( - \frac{1}{2\lambda} b^2
\,+\, b \cdot \partial_\mu \theta_\mu )  
\end{equation}
and the corresponding ghost action becomes
\begin{equation}\label{eq:defsgh}
S_{gh}[{\bar c},c;\theta]\,=\, \int d^3x \left[
\partial_\mu {\bar c} \cdot D_\mu^L c \,+\,
\frac{1}{2g} (\partial_\mu {\bar c} \times c)^a 
G_{\mu\nu}^{ab}(\theta)  (\partial_\nu{\bar c}\times c)^b \right]\;.
\end{equation}
The existence of a quartic term in the ghosts makes it evident that the
BRST transformations are not of the standard form. Indeed, we find
that the precise form for the transformations is:
$$
\delta \theta_\mu^a \;=\; \xi \, (D_\mu c)^a \,+\, \xi \frac{i}{g} f^{abe}
G_{\mu\nu}^{bd} (\partial_\nu {\bar c} \times c)^d c^e 
$$
\begin{equation}\label{eq:defbrst}
\delta c \,=\, 0 \;,\;\;\;
\delta {\bar c} \,=\, i \xi b  \;,\;\;\; 
\delta b \,=\, 0 \;. 
\end{equation} 
They leave the action $S$ invariant, and the transformation is besides 
nilpotent.

The generating functional for the gauge-fixed action is then defined 
as follows:
$$
{\mathcal Z}[J;j, {\bar\eta}, \eta] \;=\; \int [{\mathcal D}\theta]
\,{\mathcal D}b \, {\mathcal D}{\bar c} \, {\mathcal D}c \;
$$
\begin{equation}\label{eq:defztheta}
\times \,\exp \left\{- S[\theta;b, {\bar c}, c] + 
\int d^3x ( J_\mu \cdot \theta_\mu +  j \cdot b \,+\,
{\bar\eta}\cdot c + {\bar c} \cdot \eta  ) \right\} \;.
\end{equation}

It should be noted that, in all of the above equations, the covariant
derivative is defined in terms of the dependent field $L$, which
is a function of $\theta$.

This may be thought of as the main result of this letter, namely,
there exists a gauge invariant description  for the non-linear
$\sigma$-model in $2+1$ dimensions; that description is built in terms 
of $\theta$, a pseudo-vector field in the algebra of the group.
The gauge algebra is however open, what makes the BRST quantization
less immediate than for the Yang-Mills case (although the algebra is
Abelian on-shell). The resulting gauge fixed action contains terms
quartic in the ghosts, and is invariant under a global BRST symmetry. 
This BRST symmetry may be applied to, for example, the derivation of
Ward identities that will restrict the form of the counterterms.

Regarding the quantum corrections, it should be noted that there is
another (equivalent) possibility to tackle the problem of open gauge algebras,
through the introduction of auxiliary field. Their function is to
render the on-shell symmetry into an off-shell one, where the
Faddeev-Popov trick may be applied. The upshot of this procedure here, 
leads one to the `polynomial formulation' Lagrangian of
(\ref{eq:defleuc}), whose renormalization properties have been
considered in~\cite{fm1}.

\section{Conclusions}\label{sec:conc}
We have shown that the $SU(N)$ non-linear $\sigma$-model in $2+1$
dimensions may indeed be described by a gauge invariant action 
$S_{inv}[\theta]$,  for a single pseudo-vector field $\theta$. That
action has a gauge invariance which involves a composite field $L$
(a function of $\theta$) that plays a role similar to a connection. 
This, however, is so only when one considers infinitesimal 
gauge transformations. Finite gauge transformations, and the
composition of two gauge transformations show that the gauge algebra
is open. The resulting classical theory shows no difference with the
standard formulation of the non-linear $\sigma$-model, since the
classical trajectories are the only important part of the action, and
there the algebra closes. 

For the quantum theory, however, the situation is more complicated, as
the BRST quantization requires the introduction of a term which is
quartic in the ghosts. However, the corresponding global BRST symmetry
exists, and may indeed be used as a starting point in the construction
of the quantum effective action. We also note that this open algebra
formulation is also equivalent to the polynomial formulation, where
the algebra is closed and Abelian.

%====================================================================
\section*{Acknowledgments:}
The authors wish to thank The Abdous Salam ICTP, where this work was
initiated, for the warm hospitality.  C.~D.~F. was supported by
Fundaci\'on Antorchas, Argentina. C.~P.~C. thanks Olivier Piguet for useful
discussions and a careful reading of the manuscript.  
%====================================================================

\end{document}